# Potentials and Challenges of Cryoseismology with Fiber Optic Sensing in the High Arctic: A pilot experiment in Hornsund, Svalbard

Wojciech Gajek[*,1] , Max Benke[2] , Andreas Wüstefeld[3] , Andreas Köhler[3,4] , Charlotte Bruland[3] , and Alfred Hanssen[4]

**Abstract**

Distributed Acoustic Sensing (DAS) has emerged as a promising tool for environmental and cryoseismological studies, yet its performance under the extreme conditions of the High Arctic remains poorly documented. Here we report on a multi-season DAS experiment conducted across tundra and glacier environments in Hornsund, Svalbard, using 9 km of fiber-optic cable. The study combines a description of the deployment strategy, instrumentation, and operational constraints with an exploratory analysis of the recorded data to assess the types of cryospheric processes that can be captured with DAS. We document logistical, environmental, and technical challenges and provides guidelines for future experiments, including issues related to coupling, noise sources, cable integrity, and seasonal accessibility. Furthermore, we demonstrate how the dataset can be used for detecting permafrost freezing using noise interferometry, locating icequakes and calving events, as well as monitoring runoff from river-induced seismic noise. The experiment provides a field-based reference for the design and interpretation of future DAS studies in Arctic environments and highlights considerations relevant for long-term cryoseismological monitoring.

1. Institute of Geophysics, Polish Academy of Sciences, Warsaw, Poland; 2. TU Bergakademie Freiberg, Germany; 3. NORSAR, Kjeller, Norway; 4. UiT The Arctic University of Norway

## Introduction

The Arctic, including the Svalbard Archipelago, is undergoing rapid environmental changes, with warming rates exceeding the global average and accelerating transformations in the cryosphere (Isaksen et al., 2022; Schuler et al., 2025). Monitoring these processes requires observational capabilities that can operate reliably under extreme environmental conditions and provide high temporal and spatial resolution. In recent years, distributed fiber-optic sensing, including Distributed Acoustic Sensing (DAS), has emerged as a powerful tool for continuous high-resolution measurements of strain rates over several tens of kilometers of fibers (Lindsey and Martin, 2021). These systems leverage standard telecommunication fibers as dense arrays with meter-scale channel spacing, offering unprecedented spatial coverage. In this paper, we present an evaluation of DAS in a high Arctic setting from the first land DAS experiment in Svalbard.

A particularly promising field of DAS application is cryoseismology (Podolskiy and Walter, 2016; Aster and Winberry, 2017), where it offers several advantages over traditional approaches (Walter et al., 2020). Cryoseismic sources such as glacier stick–slip motion (Köpfli et al., 2022), crevassing (Nanni et al., 2022), calving events (Köhler et al., 2016; Gajek et al., 2017), frost quakes (Romeyn et al., 2021, 2022), meltwater tremors (Walter et al., 2015), and snow avalanches (e.g., Turquet et al., 2024; Kleine et al., 2025) generate a multitude of seismic signals, suitable for monitoring and improving understanding of processes in the cryosphere. Traditional sensor networks with limited number of instruments may fail to capture or localize these events due to sparse and limited spatial station coverage. DAS provides dense and spatially coherent recordings, enabling potential detection and enhanced localization of small-magnitude events (Fichtner et al., 2025). However, DAS data are exposed to higher noise levels and only provide axial wave-field component (Lindsey and Martin, 2021). While DAS cables deployed on snow, firn, ice, or frozen ground facilitate observations of cryogenic processes, lack of sufficient coupling may limit the distance at which weak signals can be observed (Hudson et al., 2021). Nevertheless, when facing these challenges considering recently established guidelines (Fichtner et al., 2026), DAS has huge potential to improve monitoring of glacier dynamics, ice-sheet stability, and permafrost integrity. Several recent field experiments (see Fichtner et al., 2026, for a complete overview) across Greenland (Fichtner et al., 2023, 2025; Gräff et al., 2025) and Antarctica (Brisbourne et al., 2021; Hudson et al., 2021) demonstrated this transformative potential of Polar fiber-optic deployments. Furthermore, several DAS deployments on Alpine environments provided valuable insight into its potential for glacier dynamics (Walter et al., 2020), permafrost (Lindner et al., 2025), and meltwater discharge monitoring (Manos et al., 2024).

Ambient noise interferometry, where virtual seismograms are obtained from cross-correlations of the noise wavefield (e.g., Larose et al., 2015), provides another powerful avenue for extracting environmental information from continuous DAS recordings. In the Arctic, there is an abundance of natural seismic noise sources that can be exploited to monitor small temporal perturbations in seismic velocity in the subsurface with dense spatial sampling. For example, recent studies have demonstrated that DAS data can reveal structure and potentially variations in permafrost (Cheng et al., 2022; Lindner et al., 2025) and firn (Zhou et al., 2022; Fichtner et al., 2023).

The high Arctic environment presents both compelling opportunities and substantial challenges for DAS applications (Fichtner et al., 2026; Gräff et al., 2025). Harsh temperatures, logistical constraints, remote sites, and dynamic interactions with snow, ice, water, and wildlife create a demanding operational context for deploying and maintaining fiber-optic instrumentation. These challenges may be further complicated in regions such as Svalbard due to strict environmental protection and regulatory constraints that prohibit standard practices like cable trenching. At the same time, these conditions amplify the scientific value of DAS: its abilities to operate without local power along the sensor and to capture glacier and permafrost dynamics, hydrological processes and ocean–ice interactions with spatial resolution that is otherwise unfeasible. In their way-leading article, Fichtner et al. (2026) discussed the huge potential fiber-optic seismology in the cryosphere by showcasing various experiments and provided practical recommendations for aspects of fieldwork such as experiment design, deployment strategies, and data storage, among others. However, the reviewed pilot studies focused mainly on short-term deployments, leaving some challenges in much-needed long-term, multi-season and large-scale passive campaigns not fully covered. Hence, recommendations on systematic assessments of deployment and maintenance strategies, data quality factors, environmental impacts, and long-term performance are still required.

This paper has two objectives. First, drawing on our multi-season experiment in Svalbard, we provide comprehensive and practical guidance towards conducting successful long-term field operations, with emphasis on Svalbard-typical challenges, that is complementary to Fichtner et al.. Second, we demonstrate scientific insights that DAS data can offer for Arctic research. In the first part, we describe the deployment environment (tundra and glacier), explain the instrumentation and acquisition parameters, discuss logistical and technical challenges encountered, including multi-season maintenance, and provide recommendations for future Arctic DAS campaigns. In the second part, we analyze the DAS data to highlight three potential applications: we detect temperature-driven velocity changes in the subsurface using noise interferometry, discuss examples of glacial seismic events, and describe river runoff-generated noise.

## The Hornsund field experiment

We carried out the study in Hornsund, Svalbard, targeting both the permafrost-affected tundra in the vicinity of the Polish Polar Station Hornsund (*PPS*) and the Hansbreen glacier system (Fig. 1). Hansbreen is a 16 km long polythermal tidewater glacier that terminates in the Hornsund fjord with a 1.5 km wide and up to 50 m high calving front (Glazovsky et al., 1991).

We instrumented the area with a 9 km long fiber-optic cable interrogated by an ASN OptoDAS unit placed in a technical building close to PPS (Fig. 1b). Due to limited interrogator availability, the data acquisition was performed in two phases: a shorter pilot phase in autumn 2023 and a longer including the onset of the melt-season in spring 2024. The cable was retrieved during summer 2024.

We used a standard telecommunication polyurethane cable, with 4 single-mode fibers in tight-buffered, gel-filled tubes. We chose a cable material with low coefficient of friction to facilitate pulling on the glacier surface. The cable was prepared in roughly 1 km sections on nine wooden drums. An armoured cable (extra fiberglass braid) was used for the initial 3 km (three drums) near PPS for protection against the wildlife followed by lightweight aramid-protected fiber further towards the glacier. Reinforced and lighter fiber sections weighted 92 and 35 kg, respectively (per km excluding 20 kg the drum). Both cables were certified rodent-resistant. Each drum had factory-provided telecom field connectors on both ends, providing basic water protection and eliminating field splicing. No special field connectors rated for harsh conditions and the long duration of our campaign were available, however, the standard ones performed well.

The fiber was supplemented with co-located geophones deployed in two areas: The first area was selected to address permafrost research. Here, 20 vertical-component 5 Hz SmartSolo geophones were installed around the 5-element permanent broadband seismic array HSPA encircled by a 1.7 km long fiber section (Fig. 1b). The second area targeted glacial seismicity and included six mini-arrays, each built from four DataCUBE loggers with 4.5 Hz triaxial geophones installed in shallow boreholes in the ice (Fig. 1c) (only during the Autumn 2023 campaign).

After initial trials, continuous DAS recording of the 2023 campaign spanning over 2300 channels (9.392 km) began at 11/9/2023 19:26 (UTC; hereafter used throughout) and ended at 21/9/2023 17:21. The spring recording ran continuously between 18/05/2024 17:29 and 10/06/2024 19:11 for 1500 channels (6.123 km), which accounted for the loss of fiber length due to irreparable breaks. We thus recorded both seasons with a gauge length of 10.27 m, channel spacing 4.08 m, and a sampling frequency of 1000 Hz.

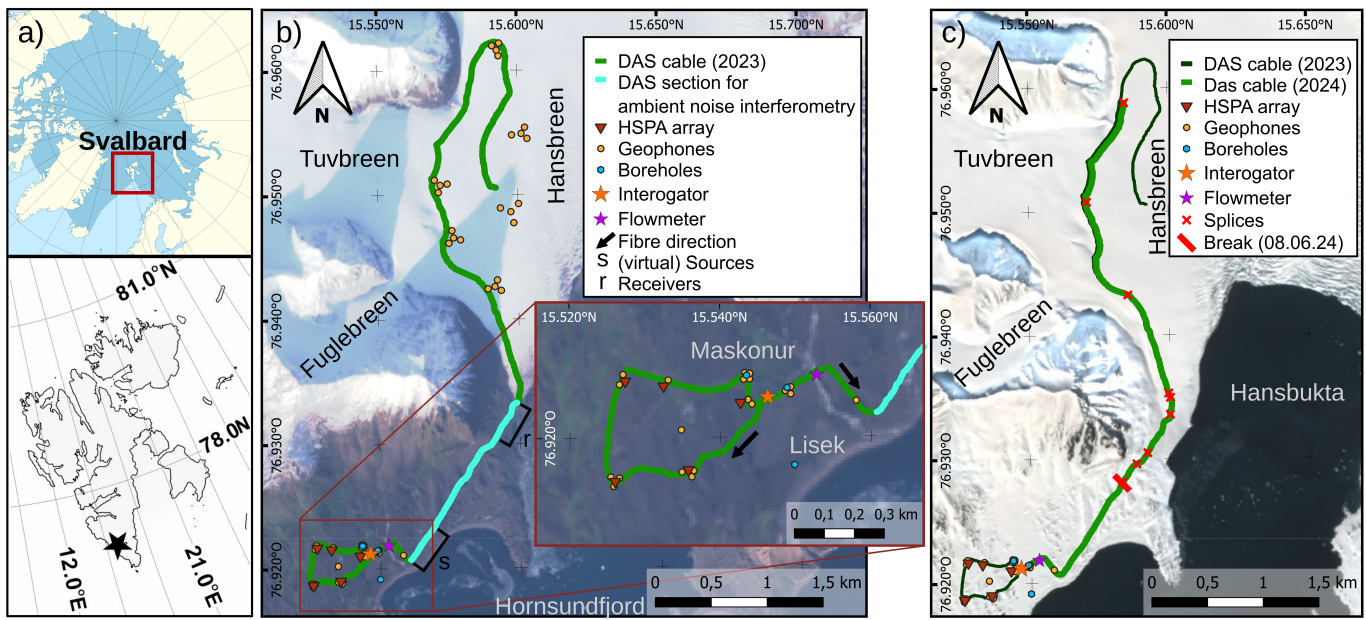

**Figure 1.** Experiment layout. (a) Top: Location of Svalbard Archipelago, bottom: Location of the Hornsundfjord (black star) in Svalbard. (b) Orthophotomap showing the geometry of the 2023 acquisition. Copernicus satellite image taken on 22/09/2023. (c) Orthophotomap showing the geometry of the 2024 acquisition. Copernicus satellite image taken on 27/05/2024.
Alt-text: Three maps showing 1) locations of Hornsund and Svalbard in the global context; 2) Autumn 2023 fiber, geophones and boreholes geometry; 3) Spring 2024 fiber layout.

## Field deployment challenges and lessons learned

Here, we describe operational aspects and associated challenges, and, drawing on our experience, discuss good practices and pitfalls that are essential to consider for executing future DAS experiments in the High Arctic.

### Instrument logistics

Local transport in Svalbard requires careful planning due to environmental regulations and season-dependent conditions. Available means of transportation are boats in the summer season, snowmobiles in the winter season, sledges on the glacier, and ATVs/tractors (in limited area). However, for final deployment, manual transport cannot be avoided. After transporting the cable drums by boat from PPS to different access points along the coast at Hornsund, we carried each drum onward in a group of four.

The interrogator is sensitive and not field-repairable. It needs to be transported with minimal vibrations in a sturdy casing. Our ASN OptoDAS unit was shipped by boat to and from PPS in 4 cruises since the unit was required elsewhere between our campaigns. The interrogator requires a reliable and stable AC power supply since power cuts may damage it. We connected the interrogator through an Uninterruptible Power Supply (UPS) unit to the PPS Hornsund power network inside a technical building located 400 m away from the station. This location offered sheltering from external conditions and noise sources, a reliable power supply, good ventilation, and remote internet access. Running a multi-week experiment on a generator in a temporary camp in Svalbard would be challenging due to the polar bear threat. Hence, we recommend research stations or hunting cabins as interrogator hosts in future experiments, even if this requires deploying longer cables.

Cable deployment strategies

We tested two principal cable-laying approaches: pulling the fiber away from a stationary drum and carrying the drum forward while unrolling. Pulling away from the drum was effective for long, straight stretches such as the glacier surface and provided better coupling due to friction-induced melt. It worked best with one person spinning the drum and required additional persons to drag the cable every 150-400 m, depending on the terrain and cable weight. The stretched group required coordinated radio communication and was more exposed to polar bear threat, requiring everyone to carry a rifle.

On the other hand, unrolling the fiber while walking and carrying the drum was more efficient for geometries including many turns or where friction from uneven terrain was high (e.g., crossing hills). This strategy kept the team together and allowed more complex cable layouts. It would be the preferred option whenever vehicles can be used (snowmobile, boat, stable sled). We found that a sturdy 2 m metal rod inserted through the drum facilitates the transportation and can serve as a rolling axle. We inserted custom spacers with rotating elements to prevent lateral drum sliding without hindering rotation (illustrated in supporting information, Fig. S5).

Sheltering the cable from wind in terrain depressions will naturally improve data quality (see Noise characterization section). However, we found that this also makes it more difficult to locate and repair the cable after the winter due to snow and ice accumulation. Following Hudson et al. (2025), one can assume bad data for poorly-coupled or free-hanging sections. Therefore, after deploying the cable, we did an inspection walk to adjust its layout and allow more slack to hanging sections. Our experience is that the cable will not self-melt into the ice from sunlight. Since the chosen black jackets offered no benefit, we recommend high-visibility colors such as orange, which are easier to locate and invisible to polar foxes.

While our fiber optic was not broken by glacial sliding-induced strain—thanks to deployment aside the main flowline where surface velocities below 20 m a$^{-1}$ were reported (Błaszczyk et al., 2024)—cryospheric deployment still exposed it to extreme strains, such as at river/meltwater channel crossings or when buried in crevasses. Avoiding heavily crevassed areas and regions with significant surface sliding velocity gradients, crossing small crevasses perpendicularly, and laying the cable along large crevasses mitigates the risk of fiber damage. Even small creeks can become powerful streams during the melt season. Plants and algae may accumulate on the cable, increasing drag. However, we found that tundra streams caused less damage than englacial channels, which turned out to be the most damaging and least predictable hazard. At the glacier terminus, meltwater relocated, tangled, and buried the cable under sediments, leading to breakage and difficult retrieval. Avoiding meltwater routes or bridging across them reduces damage risk.

Finally, georeferenced tap tests are necessary to assign channels to physical locations. We used GPS tracks to accurately interpolate between taps performed averagely every 100 m. While doing tap tests, detailed field notes and photos along the cable to document streams, moulins, hilltops, etc are essential.

For the future, we recommend using shorter (e.g., 500 m) cable sections with field connectors. This reduces the weight, allowing one person to carry the cable drum alone in a dedicated backpack, allowing for quick replacement of whole sections and for independent section health inspection with a hand-held Optical Time-Domain Reflectometry (OTDR) device, even if preceding sections are damaged. Frequent marking of the cable path with tall poles, especially in easily accessible places or at the turns, is essential since snow and ice cover make locating the fiber difficult, even with existing GPS tracks.

Cable repair and field splicing

When restarting the experiment in spring 2024, we gradually discovered seven cable breaks (Fig. 1). The loop at HSPA was cut off, as the fibre was broken and relocated by reindeer. The remaining breaks were caused by polar foxes. We performed eight field splices in total. Locating and excavating the damaged sections was time-consuming, whilst numerous splices led to signal quality losses. Therefore, we provide a dedicated lessons-learned section here.

A lightly covered fiber can be easily recovered to locate the damage. However, finding a damaged section under deep snow using only the GPS coordinates is difficult, especially if the cable lies within an ice layer, and requires laborious trenching along the cable using spades, drills, and pickaxes. Gauge length and channel spacing smears jumping or snowmobile driving signals, and, together with radio communication delay, most often prevent finding the exact damaged section. As excavating along the cable from a healthy section is too laborious and poses damage risk, we learned that installing a bypass between two known points (before and after the break) works the best. Note that two splices will be required anyway if there is no slack at the damage location.

Field splicing in freezing conditions is difficult and requires preparation. The fibre and coating become brittle. Splicing various cables at negative temperatures before fieldwork can help on cable selection. For cable repair, we used long bypass sections or reconnected on the spot. Whenever possible, the ends should be prepared under favourable conditions: ends must be peeled with the shrink tube attached, cable inserted into the waterproof splice box and possibly other protection. Doing this facilitates half of the work, but the rest must be done in the field. From our experience, the cable needs to be excavated from snow and ice at least 50 cm, preferably 1 m. If deeply buried, a snow pit needs to be dug large enough for the operator to sit by the splicing box and an assistant. Digging in shelters against the wind while a solid entrance increases personnel and fresh splice safety. A camping mat provides insulation and catches cable scraps. The operator's hands will get cold, however, wearing thin gloves still allows to work except for a few crucial moments. After successful splicing, heating may be applied twice to ensure that the shrink tube melts properly. Before closing the waterproof splice box, one should ensure the good quality of the connection. We found that the splicer signal loss estimation cannot be trusted. The only reliable option is an immediate Optical Time-Domain Reflectometry (OTDR) check on the interrogator, confirmed via radio or other communication means. Finally, the splice box should be sealed. It is best to cover the replaced section with fresh snow or trench into the snow using a chainsaw or dedicated ploughing device (Klaasen et al., 2021).

Overall, we found that field splicing under freezing conditions is possible with appropriate preparation, but poses a significant quality loss risk that is difficult to mitigate. We noted signal loss between 0 and 6 dB at spliced sections, with an average of 3.5 dB (visible in Fig. 2c). In some cases, redoing the splice did not improve the signal quality, indicating another potential damage located very close to the splice that is difficult to identify without full access to the fibre, due to the gauge length. Reducing the gauge length during repairing may help. To minimize splicing, we recommend using short by-pass sections (e.g., 500 m) with pre-installed connectors enabling rapid connecting and exchanging whole damaged sections. Well-marked field connectors enable sectional quality check with a hand-held OTDR device regardless of cable continuity.

### Wildlife threat

Wildlife is a major threat to cable integrity: reindeer and potentially polar bears, but especially curious arctic foxes. Reindeer may get entangled in the fiber with their antlers when breaking ice to feed and drag the cable. In our case, the loop section was moved tens of meters and finally broken, letting the reindeer free. This risk is limited to shallow-snow periods but difficult to mitigate. Polar bears primarily endanger personnel rather than the cable. When splicing in a snowpit, it is essential to have a guard. Arctic foxes are able to chew through reinforced rodent-protected jackets as long as not covered by snow. We did not use repellents which may be worth testing. To improve water protection and winter access on the ice, we raised the cable connectors on 150 cm poles and enclosed them in PET covers. However, it turned out that these attracted foxes, causing multiple cable damages. We therefore recommend deploying the fibre shortly before expected snowfall and not elevating any essential parts. A better alternative is a ground-level waterproof box (e.g., Pelicase) buried under snow and clearly marked.

## Potential for environmental monitoring

Despite of the challenges discussed above, we were able to successfully acquire a more than four-week-long DAS record covering different seasons. We will first characterize the data set, and subsequently provide analysis results for different applications covering permafrost, glacier, and river runoff monitoring. Note that we do not provide a full analysis of all our data here, but aim to highlight the potential of DAS cryoseismology in the Arctic.

### Noise characterization

Figure 2a illustrates the noise-level evolution along the cable during autumn 2023, with not yet connected/interrogated sections shown in dark red and the glacier edge marked by a clear, time-persistent noise contrast. A narrow zone of high amplitudes near 1.9 km corresponds to noise generated by a river, which diminishes over time as temperatures drop and the river freezes (we come back to this later). While no direct correlation with air temperature is evident, there is a clear temporal correlation of noise levels with wind speed. Since the cable is exposed, a substantial amount of the recorded noise does not originate from wind-generated seismic waves, but from wind coupling directly with the cable. This vibration can largely disturb the data analysis, in particular for ambient noise interferometry (see next section), and must be considered when processing the data. In the absence of wind, a higher noise level is recorded on the glacier section (supporting information, Fig. S4).

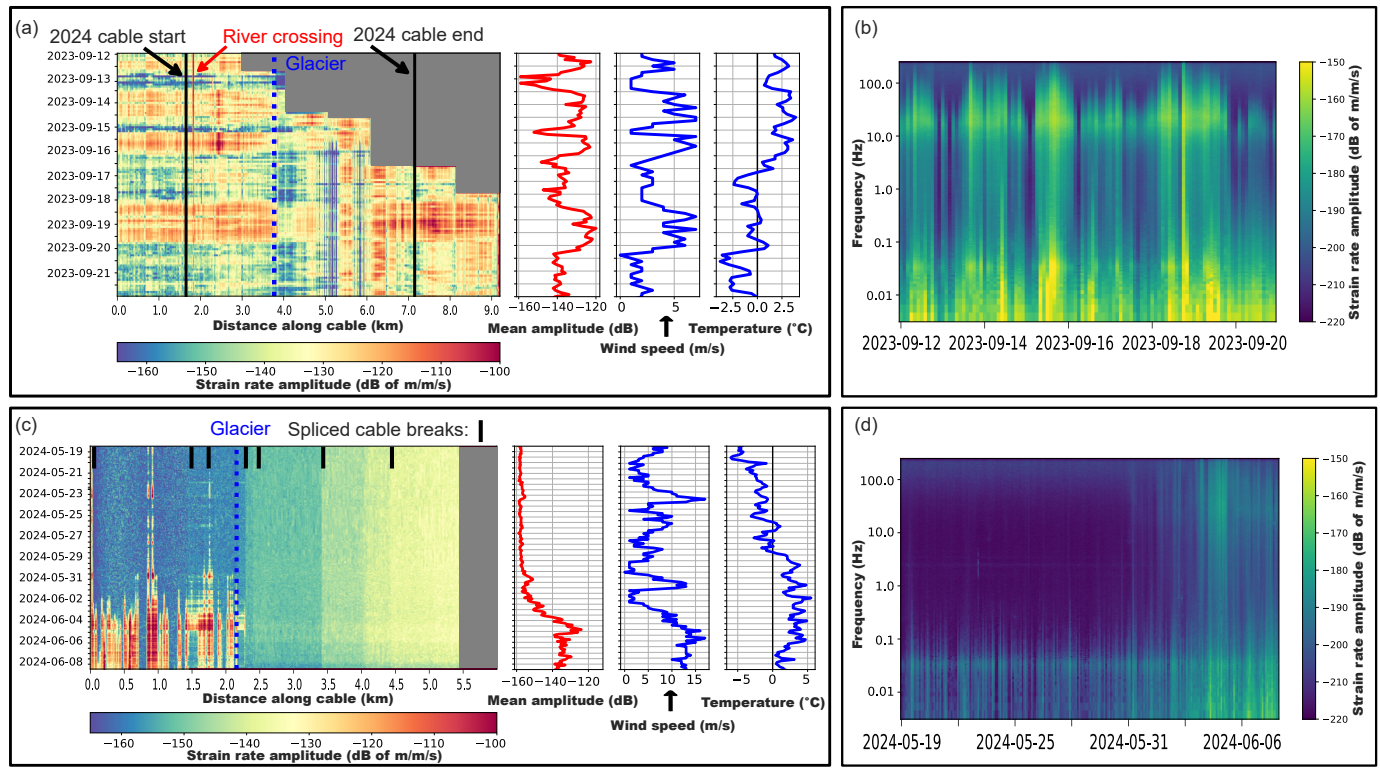

**Figure 2.** (a) Temporal evolution of the noise level along the fiber in 2023 together with the mean noise level over all channels, wind speed and air temperatures. The two vertical black lines mark the range of the data available in 2024 (c). The blue dotted line represents the edge of the glacier. A thin red line marks a small river crossing (visible high amplitudes). Unconnected cable sections are shown as gray areas. (b) Amplitude spectrogram of the spatially averaged DAS data in 2023. (c) Same as (a) but for spring 2024 data, black bars indicate splices. (d) Same as (b) but for spring 2024 data. In 2023, increase in noise levels correlates with elevated wind speeds, visible in the spectrogram around 10 Hz. 2024 data shows no wind influence. The step-wise noise increase over the distance is due to fiber damages. Temporal noise variations are due to melt season effects.
Alt-text: Spatiotemporal DAS noise analysis with associated spectrograms and meteorological data for both seasons. 2023 panels show a significant influence of wind aligned with times of elevated wind speeds that is visible around 10 Hz band on the spectrogram, and a change in noise levels when fiber enters the glacier. 2024 panels show a lack of wind influence, and a step-wise noise increase over the distance due to fiber damages and temporal noise variations due to melting snow.

In 2024, the main feature is the increase in noise levels after the onset of the melt season that begins in June (Fig. 2c). Nearly all channels were covered with snow until then, except for a small section with high amplitudes around 0.9 and 1.8 km exposed to wind due to topography. The amplitude increase is again likely related to the exposure of the cable to wind noise after snow melt. A few channels do not show this increase, most probably due to persistent snow cover. On the glacier, snow cover persists, However, noise levels exhibit abrupt step increases as a result of signal quality loss at spliced cable breaks and partially damaged cable sections. (e.g. at 4 km). At 5.5 km the cable terminates.

Next, we analyze the spectral content and its temporal variation (Fig. 2b and d). We compute spectra (Power Spectral Density (PSD) using FFT) from five minutes of data per channel every second hour (up to the glacier edge) and average them spatially. Aside from saturated very low frequencies, a temporally variable band of amplified energy around 10 Hz is observed in 2023 (Fig. 2b), with amplitude increases correlating with wind speed (same increase as seen in Fig. 2a). Comparison with the 2024 data (Fig. 2d), when the cable was largely snow-covered, shows that this band is absent until snow melt, demonstrating that the 10 Hz amplification arises from direct wind–cable interaction rather than propagating seismic waves.

To analyze spatial spectral variations, we averaged the spectra over the whole survey period for each channel (supporting information, Fig. S1). Dense horizontal lines in the channel spectra, best visible within the frequency band of 0.1 to 1 Hz, are observed, which is the typical band of Common Mode Noise (CMN) for this interrogator. CMN refers to signal components that are nearly identical across many or all channels along the fiber (i.e., no offset), arising from sources like laser phase noise, interrogator drift, or uniform environmental perturbations. The amplified band at 10 Hz is also visible and, additionally, spatial variation most likely caused by changing wind intensities, exposure, and ground coupling along the cable. Between ca. 2.4 and 3.1 km offset, the cable crosses a moraine field, causing increasing amplitudes and a frequency shift. Additional PSD plots for selected channels are provided in the supporting information (Fig. S4).

To characterize the spatio-temporal distribution of noise sources in 2023, we analyze continuous data from the co-located HSPA broadband seismic array (Fig. 1b) using frequency–wavenumber (F-K) analysis in the 3–5 and 2–5 Hz bands with 30 s moving windows. These frequency bands lay within the optimal band for the HSPA array geometry and their choice is further motivated by later findings. The resulting back azimuths (Fig. 3) show that most ambient noise arrives from 0°–70°, corresponding to the direction of glaciated areas, dominantly Hansbreen, while an additional, stronger source only visible in the 2–5 Hz band appears intermittently from the southwest and weakens toward the end of the survey. As direct wind effects on the sheltered HSPA seismometers can be excluded, this southwestern source is likely related to ocean waves in the fjord.

Permafrost monitoring with noise interferometry

Global warming in the Arctic leads to permafrost degradation, which causes erosion, abrupt mass movements, and the release of the stored carbon into the atmosphere. Hence, improving the understanding of the permafrost systems through novel monitoring methods is an important part of mitigating geohazards and predicting accelerated global warming. The measurement of subsurface seismic velocity variations due to freezing and thawing is facilitated by the high velocity contrast between liquid water and ice (Zimmerman and King, 1986). Seasonal variations in seismic P-wave velocities of over 40% due to permafrost active layer dynamics have been observed previously by active seismic surveys in Hornsund (Majdański et al., 2022; Marciniak et al., 2022). Rayleigh waves, which often dominate noise cross-correlations, have also revealed strong velocity changes of 1% to 7%, solely by centimeters deep freezing of the soil (Bruland et al., 2025; Steinmann et al., 2021).

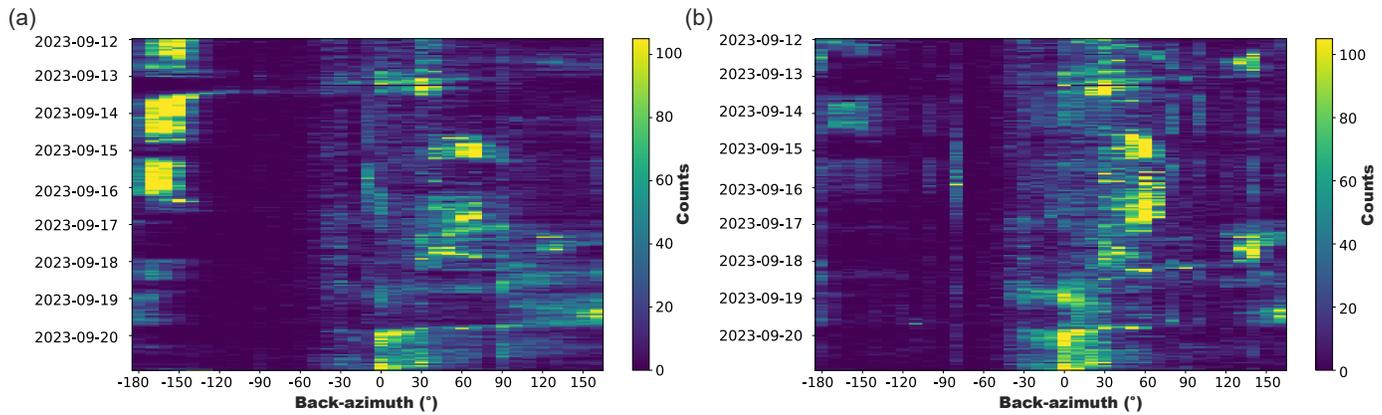

**Figure 3.** F-K analysis results of ambient seismic noise at HSPA array in the bandwidth of 2–5 Hz (a) and 3–5 Hz (b). Histograms of dominant back-azimuth values measured from moving 30 s windows are shown for two frequency bands. The 2–5 Hz band shows noise from more unstable source directions: one towards the glacier and another in the South which disappears 16/09/2023. The 3–5 Hz band shows more stable noise sources from the glacier direction (0–60°).
Alt-text: Results of F-K ambient noise directivity analysis of 2023 data. The 2–5 Hz band shows noise from more unstable sources towards the glacier and another in the South to disappear till 16/09/2023. The 3–5 Hz band shows more stable noise sources from the glacier direction.

For permafrost-focused noise interferometry in this study, we restrict our analysis to a tundra section of the 2023 dataset, during which air temperatures started to fall below zero degrees Celsius, marking the beginning of that year's freezing period. During analysis, we faced typical challenges of DAS records, such as CMN and high levels of wind-related noise recorded on badly coupled and exposed fiber sections. Obtaining reliable velocity variations from ambient noise cross-correlation functions (CCFs) requires stacking over a long time period and approximately stable noise sources in terms of location and frequency content.

*Processing*

We provide a short overview of the processing here and refer to a detailed description, including a justification for the DAS channel selection, in the supporting information (Text S1 and Fig. S3). Based on the results of the spectral analysis of the DAS data and F-K analysis at HSPA presented above, we performed cross-correlations in the 3-5 Hz band to avoid wind noise. In order to use stable source directions, we used only the cable section oriented toward Hansbreen with homogeneous frequency content over time (indicated in Fig. 1b). Data were preprocessed following Bensen et al. (2007) and using adapted code published on SeismoLive (Krischer et al., 2018). Subsequently, CCFs were computed for 30-minute windows using phase-weighted stacking. The resulting sub-stacks were further enhanced using cluster stacking (Yang et al., 2023), combined with targeted suppression of residual CMN near zero lag and manual removal of low-quality sub-stacks. CMN noise occasionally persisted, masking signals from short offsets, thus, it was necessary to only use channel pairs with a sufficiently large spacing (at least three times the expected seismic wavelength in this case) to allow selecting the most favorable sections for noise interferometry. To achieve sufficient temporal resolution under noisy DAS conditions, the final correlations relied on a combination of temporally overlapping stacks and spatial stacking of comparable channel pairs, ultimately enabling a clear and interpretable evolution of the CCFs.

*Observed time shifts*

The "Lisek" borehole at Hornsund (Fig. 1) provides subsurface temperature measurements and, due to its proximity and coastline alignment, best represents the subsurface conditions along the selected DAS section. Figure 4a shows a time-lapse plot of the retrieved CCFs together with air temperature and borehole subsurface temperature. Due to strong noise-source directionality, the CCFs are highly one-sided, and only the causal part is interpreted. Time-varying onsets occur at lag times of ∼1.1 s at the beginning of the measurement period (12/09/2023). An overall decrease in lag time is observed, which is interpreted as the effect of subsurface freezing on seismic velocities. Because overlapping time windows are used in the stacking, only the earliest stack represents fully unfrozen conditions, while later, time-shifted stacks increasingly mix unfrozen and frozen ground, producing a gradual reduction in onset lag time. The lag times stabilize after 15/09/2023, likely caused by the behavior of the stacking algorithm. A sudden shift in onset lag time becomes apparent on 17/09/2023. Afterwards, onset times stabilize at shorter lag times, indicating persistently frozen ground, while a slight additional decrease suggests continued spatial expansion of freezing. Due to limited temporal resolution and/or heterogeneous freezing along the cable, the seismic onset shift occurs 18 h after borehole temperatures and ∼24 h after air temperatures first drop below zero. Nevertheless, the results clearly demonstrate that subsurface freezing can be successfully tracked using DAS combined with ambient noise interferometry.

*Quantification of velocity changes*

To quantify the velocity increase, additional temporal stacking of the CCFs is applied, producing three overlapping stacks (72 h time shift) that represent fully unfrozen conditions, the freezing phase, and fully frozen conditions (Fig. 4b). Closer inspection of the CCFs reveals two distinct onsets at ∼1 s and ∼2 s, potentially corresponding to different wave types. Using the inter-receiver distance of 1517 m, velocities are estimated from these onset times under the different subsurface conditions. Both manual onset picking and automatic time-shift estimation, based on auto-correlation of short CCF windows at different lag times, are applied. The results are shown as onset picks and as automatically derived time shifts in Figure 4b and c.

For the first onset, we obtain a velocity of 1517 m/s. This corresponds to P-wave velocity of ∼1500 m/s for unfrozen subsurface conditions near the shore reported by Majdański et al. (2022), who investigated seismic velocities under unfrozen and frozen conditions along an active-seismic profile in Hornsund, perpendicular to our DAS profile. Hence, we interpret the first onset in the CCF as a direct P-wave. Under frozen conditions, we observe a 39% velocity increase based on the measured time shifts, yielding a velocity of ∼2100 m/s. This value falls within the 2000–3500 m/s range reported by Majdański et al. (2022) for frozen ground. Our estimate close to the lower bound suggests that the subsurface is not fully frozen. This is consistent with borehole temperatures indicating that freezing is limited to the uppermost meter and that P-waves retrieved from noise are expected to sample greater depths.

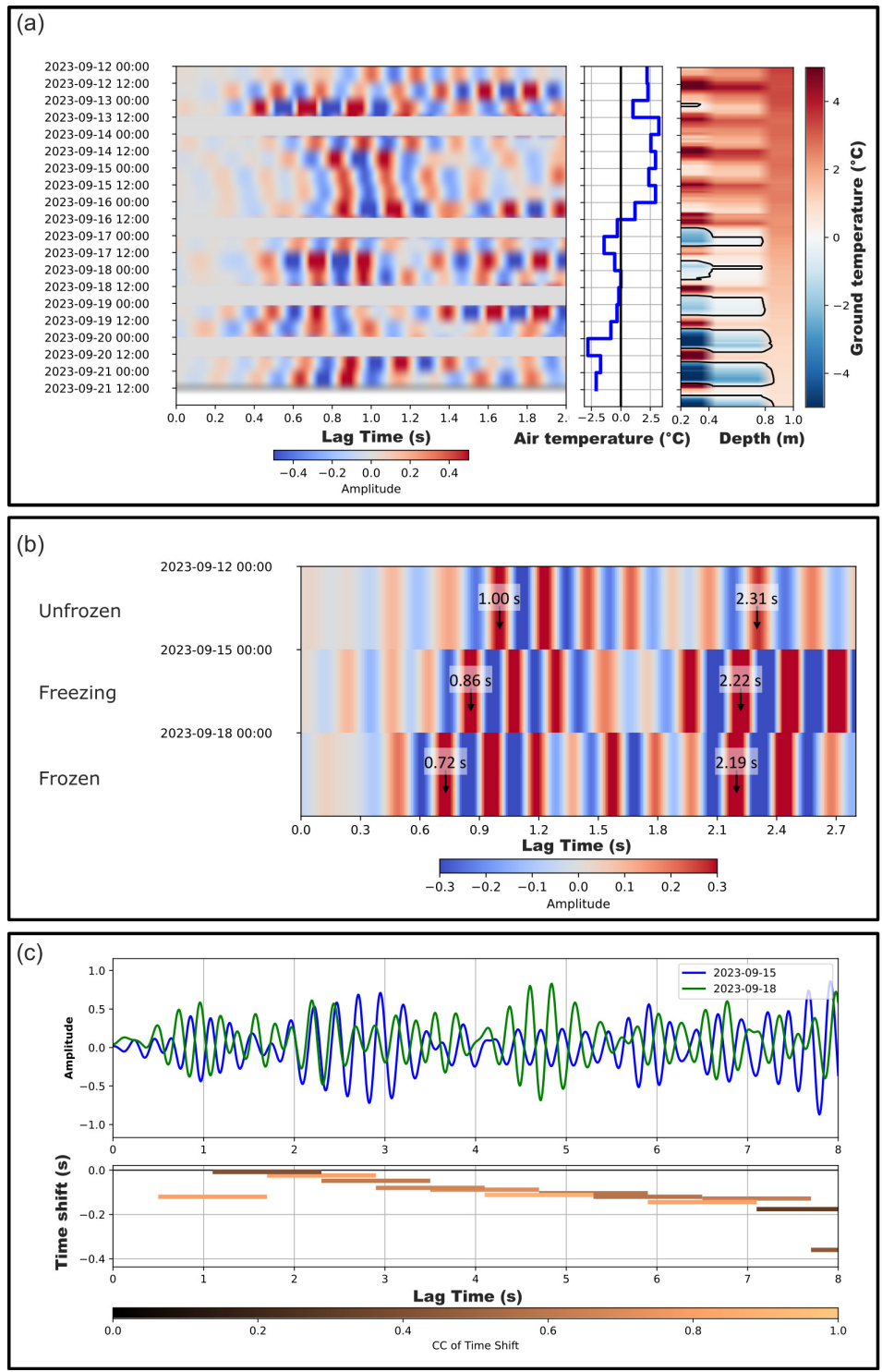

**Figure 4.** DAS noise cross-correlation results. (a) CCFs obtained from stacks over 120 h with 12 h overlap, spatially stacked over 50 channels. Frequency band is 3–5 Hz. Low-quality stacks are masked. The same time scale is used for air and ground temperatures at "Lisek" borehole to the right. (b) CCFs from (a) are further stacked over three time intervals of different stages of ground freezing to further enhance the signal-to-noise ratio. Manually picked arrival times are shown. (c) Waveforms of first and last CCF from (b) comparing unfrozen to frozen time intervals (top) and automatically measured corresponding time shifts of the CCFs at different lag times (bottom).
Alt-text: 2023 CCFs with phase shift aligned in time with air temperature drop below 0 and episodic freezing of the top first m of soil in the "Lisek" borehole. This is followed 120-h resolution CCFs with picked arrivals of P- and surface waves that shift in time due to freezing soil. Finally, waveforms of the first and last CCF show gradually increasing phase shifts in the coda.

For the second onset, velocities increase by ∼5%, from ∼660 m/s to ∼690 m/s. This onset is interpreted as a surface wave, most likely a Rayleigh wave. Given the dominant noise source alignment with the cable and the directional sensitivity of DAS, the emergence of S or Love waves in the CCFs is not expected. Signals following the direct surface-wave onset are interpreted as coda waves. Since they consist of scattered energy traveling longer paths and sampling a larger area, they are more sensitive to velocity changes. Consistently, we observe progressively increasing time shifts within the coda from unfrozen to frozen conditions (Fig. 4c). Coda waves may also sample spatially heterogeneous freezing conditions over a broader area surrounding the cable.

The F-K velocities derived from the vertical components of the HSPA seismometer array (supporting information, Fig. S2) exhibit clear dispersion, indicating the presence of surface waves. For most of the survey period, velocities remain near 2200 m/s, but on the evening of 19/09/2023, nearly coincident with the onset of freezing at that site, they rapidly increase to ∼2600 m/s. While this provides independent evidence for a temperature-driven seismic velocity change, both the absolute values and the relative increase are significantly larger than those obtained for the surface wave signals inferred from DAS. This suggests that the F-K estimates may be influenced by a mixture of P and Rayleigh waves, highlighting the advantage of noise interferometry in separating different wave types.

Glacial seismic activity

We present DAS records of a strong icequake and calving signals that occurred in September 2023 and compare them with vertical component geophone data. Importantly, the geophones were drilled 0.5 m into the ice and thus the noise level is lower than in the DAS data, as the fiber was deployed on the ice surface.

We can clearly see the icequake signal on all geophones, whilst in DAS data we could only observe it up to 1 km away from the source location (Fig. 5). The geophone records allow to analyze broader signal spectrum down to 5 Hz, whilst in DAS data below 80 Hz noise interferes with the icequake signal. Nevertheless, due to superb spatial resolution, DAS data allow for observing finer details of the wavefield, including several weaker repeating icequakes that follow the main event and that are not observable in the geophone data. However, this is limited to only the strong icequakes, while weak events (well visible in geophone data) are not observed. Overall, the sensitivity of DAS (with exposed fiber) is lower than that of the geophones, while the spatial resolution is higher. On the other hand, geophones offer better signal-to-noise ratio, allowing for detecting far (>1 km) icequakes with lower spatial resolution. We cannot compare performance of buried DAS against geophones as those were not deployed during the spring. Ultimately, we learned that using both methods simultaneously provides the best efficiency, as geophone data can be used for detecting events and analyzing far field, while DAS data can enhance the location accuracy of icequakes within 1 km range from the fiber or provide additional information about glacier internal structure thanks to finer details visible in the recorded wavefield. When using DAS alone under the same conditions, very weak and distant icequakes would be overlooked, thus we see DAS as a method auxiliary to geophones.

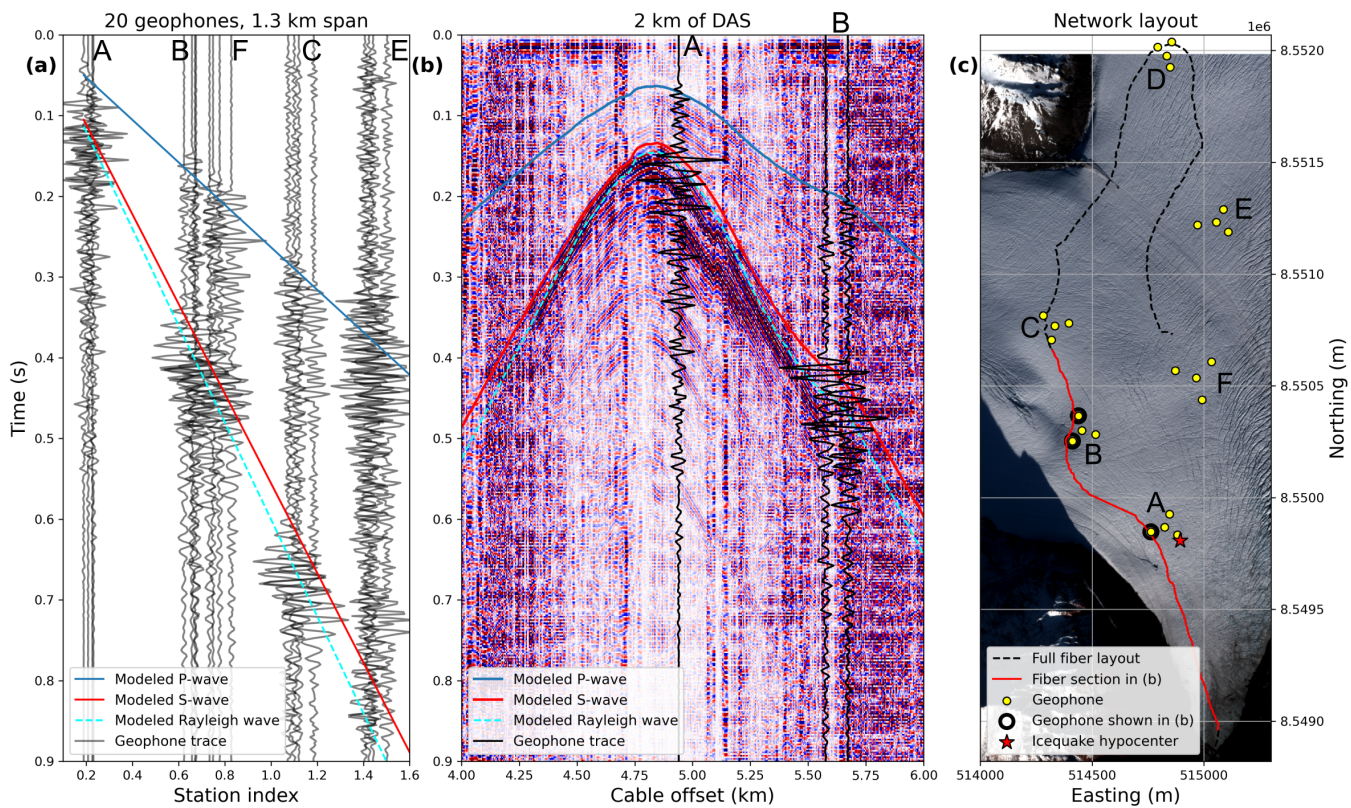

**Figure 5.** Surface icequake signal recorded on geophones and DAS. (a) Vertical ground velocity (80-150 Hz bandpass filtered and normalized to each trace maximum) recorded on 20 geophones drilled in the ice. Capital letters on top show the location of each mini-array in panel (c) - mini-array D is not shown. Colored lines show predicted movoeouts of P-, S-, and Rayleigh waves, respectively (same in (b)). Note possible refracted P-wave arrivals on the E-array. (b) DAS data section: strain rate, 80-150 Hz bandpass filtered and normalized to each trace maximum. Black lines show data from three geophones installed exactly on the fiber path, marked with a thick black circle in (c). The icequake signal was not visible on further offsets. (c) Orthophotomap showing cable and geophones layouts and the estimated hypocenter of the icequake. The estimated source depth is 130 m, which is similar to the ice thickness. Satellite photo taken on 22/9/2023.
Alt-text: Comparison of the icequake recorded on 20 geophones vs 2 km of DAS with marked modeled P-, S-, and Rayleigh wave arrivals fitting the waveforms. In DAS data, some repeaters are visible.

Calving events are usually more energetic than icequakes with more complex, low-frequency waveforms characterized by strong S-wave and very weak P-wave amplitudes. Calvings are visible in the DAS data over a wide range of offsets with varying amplitudes depending on the cable's orientation relative to the source and wind exposure. Due to waveform complexity and lack of clear onsets, calvings are challenging to accurately locate using sparse geophones. We estimate the location of confirmed calving event using the method of Baird et al. (2025) that spatially stacks the DAS waveforms using arrival times predicted for each candidate source location. Due to the high spatial sampling of DAS, this method results in a robust calving location with the highest stack value falling 150 m away from the calving front (Fig. 6), when using a homogeneous velocity of 1775 m/s that maximizes the stack value. As this value falls between the surface wave and the S-wave velocity, a more accurate velocity model that accounts for ice and bedrock should be used in future analysis to increase accuracy of source location. This example shows that land-deployed DAS can locate calving episodes, however, underwater fiber deployment is likely better suited for studying calving Gräff et al. (2025).

The presented icequake and calving examples demonstrate the potential of DAS for detecting and locating glacial seismicity and potential gains offered by dense spatial sampling. Although quantification and detailed analysis of glacial activity remain key future goals, it will depend on improving signal quality. In particular, advanced denoising and stacking strategies are essential to enhance signal-to-noise ratio, increase detection sensitivity, and enable more reliable event detection and characterization.

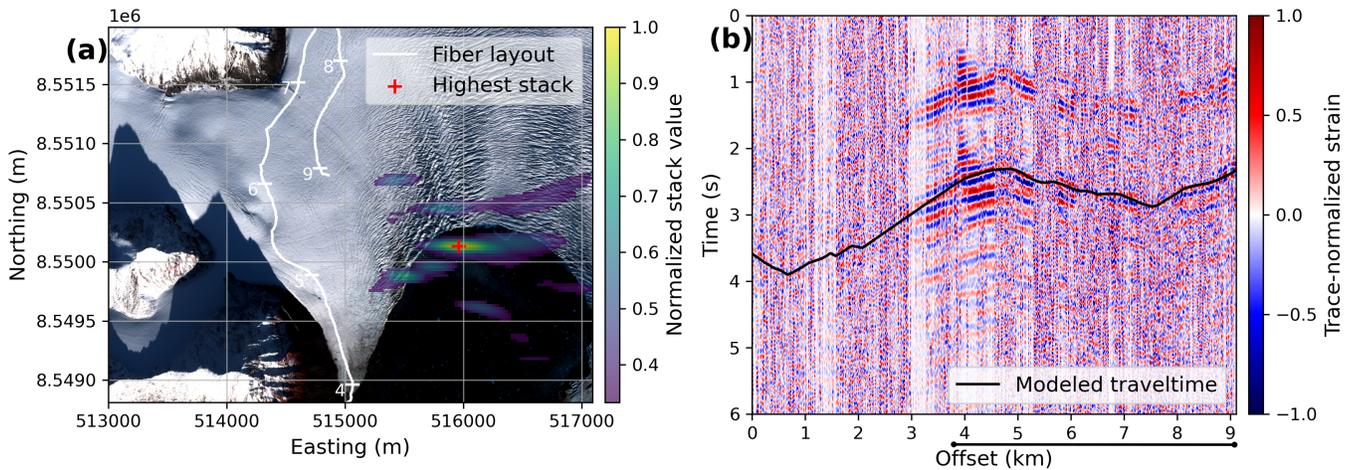

**Figure 6.** Location of calving event. (a) Orthophotomap presenting the fiber layout (white line with distance bars in km) and stack value distribution. Satellite photo taken on 22/9/2023. (b) DAS strain data bandpass filtered 3-10 Hz containing a double calving event recorded on 21/9/2023. The black line presents modeled traveltimes. The location is estimated based on the highest amplitude stack value along the expected moveout. Black arrow at the bottom marks offsets presented in panel (a).
Alt-text: Map of glacier and fiber layout overprinted with resolved location of calving event, with maximum closely aligned with the calving front position. Another panel presents the DAS data section with a visible double calving event waveform and modeled traveltimes fitting the event's moveout.

River runoff

We also use DAS to monitor river flow dynamics. We perform our analysis using the most energetic channel along a short, straight fiber section crossing the above-mentioned stream (Fuglebekken). We compare DAS-derived signal with hydrological measurements from a co-located flowmeter.

In Fig. 7, we analyze the DAS broadband signal spectrogram (Welch PSD) and DAS high-frequency (>0.3 Hz) strain amplitude RMS to track the temporal evolution of flow-induced strain. The hydrological data shows a gradual decrease in water flow followed by repeated nighttime freeze–thaw cycles until sensor removal on 19/9/2023. Active flow periods are marked by high strain amplitudes and increased energy near 10 Hz. After 16/9/2023, the strain shows strong diurnal modulation that aligns with in-situ measured discharge minima, particularly in the <20 Hz band, which is most sensitive to turbulent flow and bedload motion (Gimbert et al., 2014).

As air temperature decreases towards the end of the observation period, the river section diurnally freezes. Consequently, the flowmeter stops providing reliable measurements (freezing terminates flow velocity but increases pressure, based on which the flowmeter incorrectly interprets increased water level) and is finally removed. In contrast, DAS gathers data from beyond point measurement (equal to gauge length), and thus, remains sensitive to residual under-ice flow and mechanical coupling between the ice cover, water and riverbed, preserving a dynamically rich signal regardless of freezing.

We compute Pearson correlation coefficients between the 30-min averaged DAS strain RMS and the hydrological data. We observe a strong positive correlation ($r = 0.84$) between strain RMS and measured flow velocity. The correlation with volumetric discharge (flux) is lower ($r = 0.66$), while the water level shows a negligible negative correlation ($r = -0.20$). These results confirm that the high-frequency DAS strain amplitude reflects dynamic flow processes (velocity and associated turbulence). This behavior highlights the potential of DAS to serve as a proxy for river discharge when direct flow measurements fail and motivates the development of DAS-applicable empirical or physics-informed transfer functions that use strain-based observables alone to reconstruct discharge time series including freeze-thaw cycles.

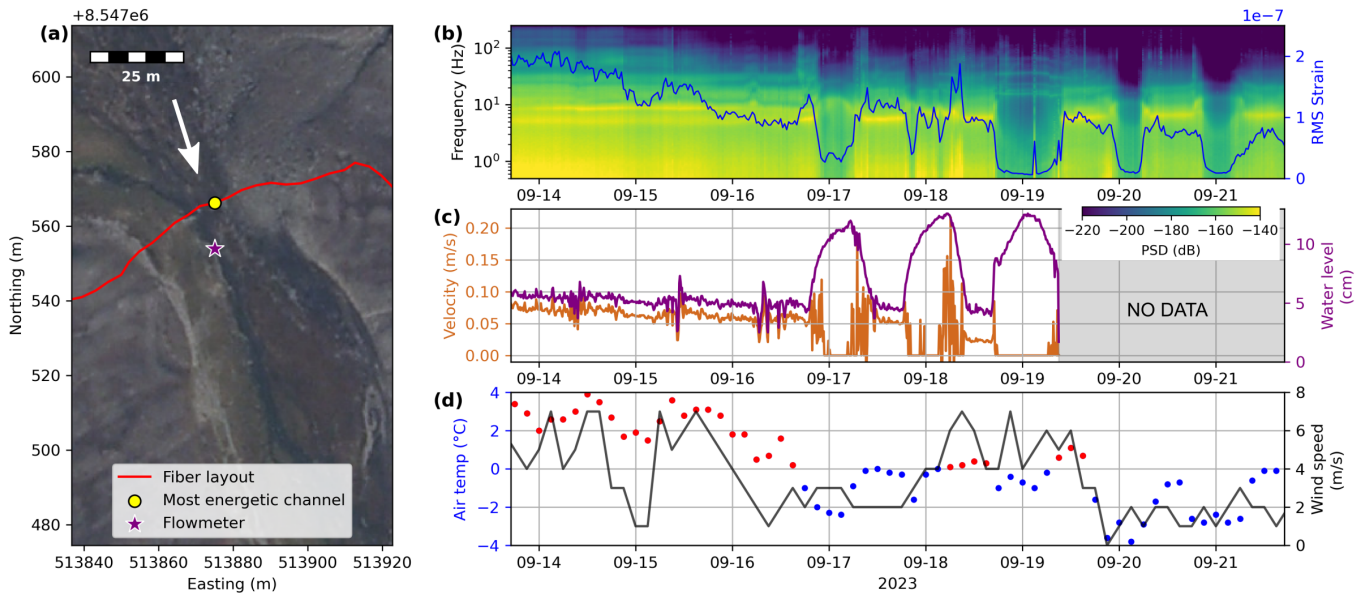

**Figure 7.** River runoff analysis. (a) Map-view showing stream at Fuglebekken on aerial imagery from 2011 and locations of the fiber (red line), the flowmeter (purple star), and main channel during measurement (white arrow). A yellow dot marks the analysed channel. (b) DAS data: spectrogram and RMS of strain (blue line). (c) Hydrological data measured with Nivus PCM-F with an active Doppler sensor at the bottom of the Fuglebekken stream at the fiber location: orange line shows water velocity and purple line shows water level. (d) Meteorological data from nearby PPS Hornsund: Red and blue dots show positive and negative air temperature, respectively, and the black line shows wind speed.
Alt-text: Spectrogram of DAS data and strain RMS aligned with hydrological (water level and velocity) and meteorological (air temperature and wind) data. Periods when the stream was frozen (no flow velocity) correlate well with the measured strain decrease and drop of energy below 20 Hz.

## Conclusions

This study presents preliminary results of the first multi-season land DAS deployment in Svalbard. Using a 9 km fiber-optic cable deployed over tundra and glacier environments near the Polish Polar Station Hornsund, we demonstrate both the scien-tific potential and the practical challenges of applying DAS to cryoseismology under extreme environmental conditions. Our results show that despite substantial logistical and technical constraints, DAS can provide spatially continuous observations of permafrost dynamics, glacial seismicity, and hydrological processes in High Arctic settings.

From an operational perspective, the experiment highlights that long-term DAS deployments in the Arctic are feasible but require careful planning and maintenance. Key challenges include cable exposure to wind-induced noise, wildlife interactions, cable damage, and the difficulty of locating and repairing cable breaks beneath snow and ice. Combining the lessons learned from cable deployment and noise analysis, we recommend future layouts including diverse directions of straight fibers that are long enough (at least three times the expected seismic wavelength) to allow selecting the most favorable sections for noise interferometry, depending on noise source distribution. Common mode noise should be reduced as much as possible with appropriate measures during interrogator installation (e.g., vibrations canceling mat/table). Furthermore, to better deal with cable breaks and easier deployment, modular cable sections with pre-installed connectors and layouts where the cable returns multiple times to the location of the interrogator (NORSAR, 2023), combined with interrogators being able to interrogate multiple fibers in parallel, are beneficial for future experiments. Dense marking of cable routes and avoiding meltwater channels are essential for improving winter survivability and facilitating post-winter recovery.

Scientifically, the data demonstrate that DAS can resolve seismic signals across a wide range of processes. Although data quality required careful preprocessing and extensive stacking, ambient-noise interferometry successfully captured the onset of seasonal ground freezing, revealing velocity increases consistent with borehole temperature measurements and previous active-seismic studies in the area. This capability is particularly promising for monitoring heterogeneous permafrost conditions that are difficult to resolve with sparse point sensors. The observations of glacial seismicity further illustrate the strengths of DAS for detecting and locating events, especially when deployed together with geophones. Similarly, the river runoff analysis demonstrates that DAS is sensitive to flow-related seismic signals even when conventional hydrological instruments fail during freeze-up, highlighting the potential of DAS as a proxy for discharge monitoring.

Overall, this pilot experiment shows that DAS can substantially enhance environmental seismology in the high Arctic. As Arctic warming accelerates, scalable monitoring approaches that minimize environmental impact while maximizing spatial coverage are urgently needed. The lessons learned and examples presented here contribute to an emerging framework for robust, long-term fiber-optic monitoring in polar regions and provide practical guidance for future DAS deployments targeting permafrost stability, glacier dynamics, and hydrological changes.

## Data and Resources

This study utilises meteorological and hydrological data from the monitoring carried out based on the infrastructure of the Stanisław Siedlecki Polish Polar Station Hornsund in Spitsbergen, operated by the Institute of Geophysics, Polish Academy of Sciences, financed by the Polish Ministry of Science and Higher Education. Borehole temperature data were obtained from the grant no. 2020/38/E/ST10/00139 funded by the Polish National Science Centre.

The geophone data gathered within the project are accessible through the IG PAS repository: https://doi.org/10.25171/InstGeoph_PAS_IGData_FROST_geophone_seismic_data_Hornsund_2023_2024. Due to its large volume (11 TB), DAS data are available on request. HSPA waveform data are available from the Norwegian EIDA node (Ottemöller et al., 2021). The Copernicus Sentinel satellite imagery associated with Fig. 1 was accessed through Copernicus Browser: https://dataspace.copernicus.eu/browser/ on 19/05/2025. The map in Fig. 1a was modified from "Arctic Ocean SVG" by Norman Einstein via Wikimedia Commons: https://commons.wikimedia.org/wiki/File:Arctic_Ocean_SVG.svg.

A video presenting Hansbreen environment and cable deployment is available at: https://www.youtube.com/watch?v=VTQ0QaDGrKk.

The Supplemental Material associated with this paper contains a detailed description of seismic noise interferometry processing, together with accompanying figures and spatial PPSD estimates of the seismic noise.

## Declaration of Competing Interests

The authors acknowledge that there are no conflicts of interest recorded.[1]

## Acknowledgments

The study was funded by the Research Council of Norway, project number 322387, Svalbard Integrated Arctic Earth Observing System - Knowledge Centre, operational phase 2022, and the National Science Centre, Poland (NCN), grant number 2024/55/D/ST10/0199. Access to Planet imagery was granted within the Svalbard Integrated Arctic Earth Observing System-–Planet cooperation project. We thank Mariusz Białecki, Ali Gholami, Andrzej Górszczyk, Przemysław Matwiejczuk, Mateusz Olszewski and Szymon Szyszko for fieldwork assist. We also thank Marzena Osuch for sharing the borehole temperature data and Alan Baird for pylops scripting.

---

[1]The authors acknowledge that there are no conflicts of interest recorded.

# Supporting Information for

## Potentials and Challenges of Cryoseismology with Fiber Optic Sensing in the High Arctic: A pilot experiment in Hornsund, Svalbard

Wojciech Gajek[1], Max Benke[2], Andreas Wüstefeld[3], Andreas Köhler[3,4], Charlotte Bruland[3], Alfred Hanssen[4]

[1]Institute of Geophysics, Polish Academy of Sciences, Warsaw, Poland
[2]TU Bergakademie Freiberg, Germany
[3]NORSAR, Kjeller, Norway
[4]UiT The Arctic University of Norway

Corresponding author: Wojciech Gajek (wgajek@igf.edu.pl)

# S1 Seismic noise interferometry processing

## S1.1 Common Mode Noise

Common Mode Noise (CMN) influence on the ambient noise cross-correlations can be severe. (Willis, 2022) presented a strategy for CMN removal that was adopted in our processing. Following Willis (2022), all channels of the data are first stacked along the spatial axis. Doing so, propagating seismic signals showing moveout (distance-dependent arrival times) are in most parts removed by stacking, while the CMN without moveout is stacked constructively. Subsequently, after division by the number of channels (i.e., computing the spatial mean), the stack is subtracted from each channel of the original data, thus removing the CMN.

For our data, an improved version of this approach was required: Instead of stacking the data to obtain the spatial mean, we subtracted the spatial median from the original data, which was more stable towards outlier channels. As channels positioned further apart are expected to show larger time differences in the seismic signals' arrivals, while the CMN remains constant, we used every n-th channel for the calculation of the median, to reduce the computational effort of CMN removal. We found a minimum of 100 channels to be optimal for CMN removal in this study. Calculating the median was only done inside the frequency band of $0.02 - 2\,\mathrm{Hz}$ (used interrogator-typical CMN band). In Fig. S1 we present the data before and after CMN removal. Note that CMN outside of its typical frequency band cannot be removed efficiently with the presented method. However, it is still possible to identify the CMN and account for it after computing the cross-correlation.

## S1.2 Frequency band selection

The spectral content of the DAS data is highly relevant for the calculation of the noise cross-correlations. Frequencies heavily influenced by wind should be avoided since these signals do not sample the subsurface. Therefore, based on the results of the noise analysis (Fig. 2b), we used an upper boundary of 5 Hz to avoid wind noise. Furthermore, sudden changes of noise source direction can have a significant effect on the cross-correlation results, leading to the obtaining of false velocity variations. Hence, considering the results of the F-K analysis (Fig. 3), we concluded that frequencies of 2 Hz and below should be avoided due to temporally unstable noise source directions. Hence, considering spectral and F-K analysis together, an optimal frequency band of 3 to 5 Hz was used for the cross-correlations.

## S1.3 Channel selection

Analysis of the spatial variability of the frequency content (Fig. S1) allowed us to select optimal channel pairs for cross-correlation. The spectra for the time windows of both channels to be cross-correlated need to be similar. The channel amplitudes are relatively homogeneous within the frequency band of 3–5 Hz in the first 4 km of the cable. Only in a section from 2.4 to 3.1 km a significant difference in the frequency content is visible. Therefore, this section was excluded.

Furthermore, the results of the F-K analysis (Fig. S2 helped us to identify optimal channels for the cross-correlations. Within the 3–5 Hz band, the noise originates mainly from Hansbreen. Considering the directional sensitivity of DAS and the primary contribution to the virtual signals of sources inline with the receivers, it is therefore favorable to use data from a fiber section that is oriented directly towards the glacier.

We observed that CMN is often not removed completely when a peak of high correlation appeared at zero lag time in the cross-correlations. This CMN signal concealed potentially useful information about velocity variations at short lag times. The problem persisted even if the CMN signal was suppressed (masked) after the correlation (see processing below). Therefore, it was necessary to only use channel pairs with a sufficiently large spacing, so that potentially useful signal onsets appear outside of the CMN window. Finally, we found that, in addition to temporal stacking, spatial stacking of the cross-correlation was necessary to retrieve satisfying results (see processing below). All channels used for the spatial stack must be as similar as possible regarding the above described conditions (i.e, frequency content and orientation). The only cable section with a sufficient number of such channels is indicated in

Fig. 1. To satisfy the condition of similar noise characteristics and large enough distance of the channels to be correlated, channels at the beginning and end of this section were used for noise interferometry in this study (marked as *s* for sources and *r* for receivers in Fig. 1).

## S1.4 Processing

### S1.4.1 Pre-processing, correlation and basic stacking

The processing workflow is summarized in Fig. S3. Pre-processing of the data included chunk-wise loading, CMN removal, downsampling, detrending, tapering, bandpass filtering, 1-bit normalization, and spectral whitening. Due to memory limitations, the data was loaded in 30-min-chunks. After temporal downsampling to 125 Hz, we removed CMN as described above. The subsets were then assembled again to form the complete dataset of almost 10 days, starting on 12/9/2023 at 00:00:10 UTC and ending on 21/9/2023 at 12:00:10 UTC. The data contains two gaps from 16/9/2023 21:00:10 UTC to 17/9/2023 00:00:10 UTC and from 18/9/2023 09:00:10 UTC to 18/9/2023 10:00:10 UTC. We then detrended the data, tapered at the edges with a 0.05% cosine taper, and filtered it in a broad frequency band from 0.7 to 30 Hz. Subsequently, the data was normalized in the time domain using a 1-bit normalization to remove the effect of strong seismic events (Bensen et al., 2007), in our case, mostly calving events from Hansbreen. Spectral whitening was applied as the final step of pre-processing.

The computation of cross-correlation was done iteratively for 30-min-long time windows moving over the complete dataset with no overlaps. We did not compute auto-correlations in this study. For each (pre-processed) 30-min window, cross-correlation functions (CCFs) were calculated for five-min-long data segments. This resulted in six CCFs per time window, which were then immediately stacked to form a single 30-min sub-stack. The calculation of CCFs for five-min segments was done to speed up processing. The length of five minutes per segment is more than sufficient to contain the desired signal for all possible receiver spacings. Assuming seismic velocities in unfrozen and frozen permafrost between 650 and 3500 m/s for different wave types (see main text) and the chosen frequency bands, wavelengths between 100 and 1000 m are expected. In terms of travel times for the direct waves, the maximum time delay would be around 2.5 s along the section of the fiber used here. A phase-weighted stacking algorithm (Schimmel and Paulssen, 1997) turned out to be optimal for the creation of the 30-min sub-stacks.

After testing several methods (linear stacking, phase-weighted stacking, selective stacking, cluster stacking) and choosing the best-performing one, we used a cluster stacking algorithm to further stack the 30-min sub-stacks (Yang et al., 2023). Here, the traces are first separated into two groups corresponding to their respective signal quality. Then, the traces within those two clusters are independently stacked. For the final stack, the two resulting CCFs, representing the respective cluster, are stacked using weights. If their correlation coefficient is larger than a similarity threshold of 0.75, the clusters are weighted by their respective relative peak amplitude, i.e., the ratio of the maximum to the root-mean-square of the traces. If the correlation coefficient of one of the CCF clusters is below the similarity threshold, then the CCF of the cluster with the higher relative peak amplitude is used alone as the final stack. This procedure played a decisive role in the generation of reliable results.

It is inherent to the method that traces with high SNR get amplified by the algorithm. However, in the case of DAS data, the traces with the highest SNR can also be those where CMN is still present after the initial CMN removal. To solve this problem, a combination of suppression of the CMN peak and selective stacking was employed. Firstly, a suppression window of 0.5 s around zero lag time was used to downscale (mask) the influence of the CMN peaks. Sub-stacks that afterwards were still negatively affecting the result were then removed manually prior to the stacking process (see below).

### S1.4.2  Advanced stacking

In our study, a resolution of at least 12 h was desirable to track the temperature-induced changes in the sub-surface. Stacking of 12 hours of data alone was thereby insufficient to retrieve stable results, that is still too low CCFs SNRs, most likely due to challenges mentioned above (CMN, wind noise, etc) This imposed a major problem that had to be overcome with an advanced stacking strategy: The commonly applied temporal stacking was expanded by temp-shift stacking (1), combined with spatial stacking (2), and selective stacking (3).

(1) Temp-Shift Stacking: The air temperature first falls below zero degrees on the afternoon of 16/9/2023, and freezing of the subsurface is observed in borehole data afterwards (17/9/2023). This divides the measurement period roughly into two halves. The crudest temporal resolution that still resolves this change, is therefore obtained with the use of two 120 h stacks, reaching from 12/9/2023 00:00:10 UTC to 17/9/2023 00:00:10 and 17/9/2023 00:00:10 to 22/9/2023 00:00:10. While a shift in travel times in the CCFs was already visible using this 120 h resolution stacks, the overall temporal resolution was not feasible to infer meaningful statements about the exact freezing time. However, the reduction of the total stack time led to unstable and unsatisfactory results. To enhance the resolution, while maintaining the principal stack-length of 120 h, the 120 h stack is calculated multiple times with start and end times shifted by 12 h for each iteration. This results in the traceability of the onsets in the CCFs at the desired (pseudo) 12-h resolution.

(2) Spatial Stacking: Compared to geophone data, DAS data is noisier but more densely

sampled spatially. This increased availability of spatial data can be used to compensate for the overall lower DAS data quality. Spatial stacking of raw data before computing cross-correlations has previously been used to improve coherence and temporal resolution (Maass et al., 2024). Here, we spatially stack cross-correlations. For the spatial stack, correlations are calculated for multiple channel pairs with similar distance and orientation. The selected temp-shift stack is produced for every intended channel pair and subsequently stacked spatially using linear stacking. For the presented results, a stack of 50 channel pairs was used. Note that small-scale spatial variations of the subsurface freezing-pattern are averaged out by the spatial stack. However, it does reduce the influence of single channels and their immediate surroundings on the result. This is desirable since the coupling and immediate subsurface conditions can differ significantly for the channel locations along the cable, resulting in potentially suboptimal conditions for some specific channels. We found that spatial stacking of cross-correlations must be performed after temporal stacking. Otherwise, the quality of the final stack deteriorates, making it more challenging to obtain a clear and interpretable onset.

(3) Selective Stacking: Since the full influence of CMN on the stack could not be removed, sub-stacks still containing CMN (therefore considered of low quality) were removed before stacking. The selection of these traces was done manually after temp-shift and spatial stacking, as their influence on the result was examined by trial-and-error. This method is subjective and non-reproducible, but immensely improved the result, ultimately leading to the emergence of a clearly traceable onset in the CCF.

To summarize, the CCFs shown in Fig. 4a (main text) were calculated as temp-shift stacks with 120 h stack length and 12 h shift, spanning over the entire measurement period. We did this for 50 channel pairs separated by ca. 1.5 km, which were subsequently spatially stacked. The channels 530 to 580 (2.12 to 2.32 km) were used as virtual sources, and channels 900 to 950 (3.6 to 3.8 km) as the associated receivers.

# S2 Supplementary Figures

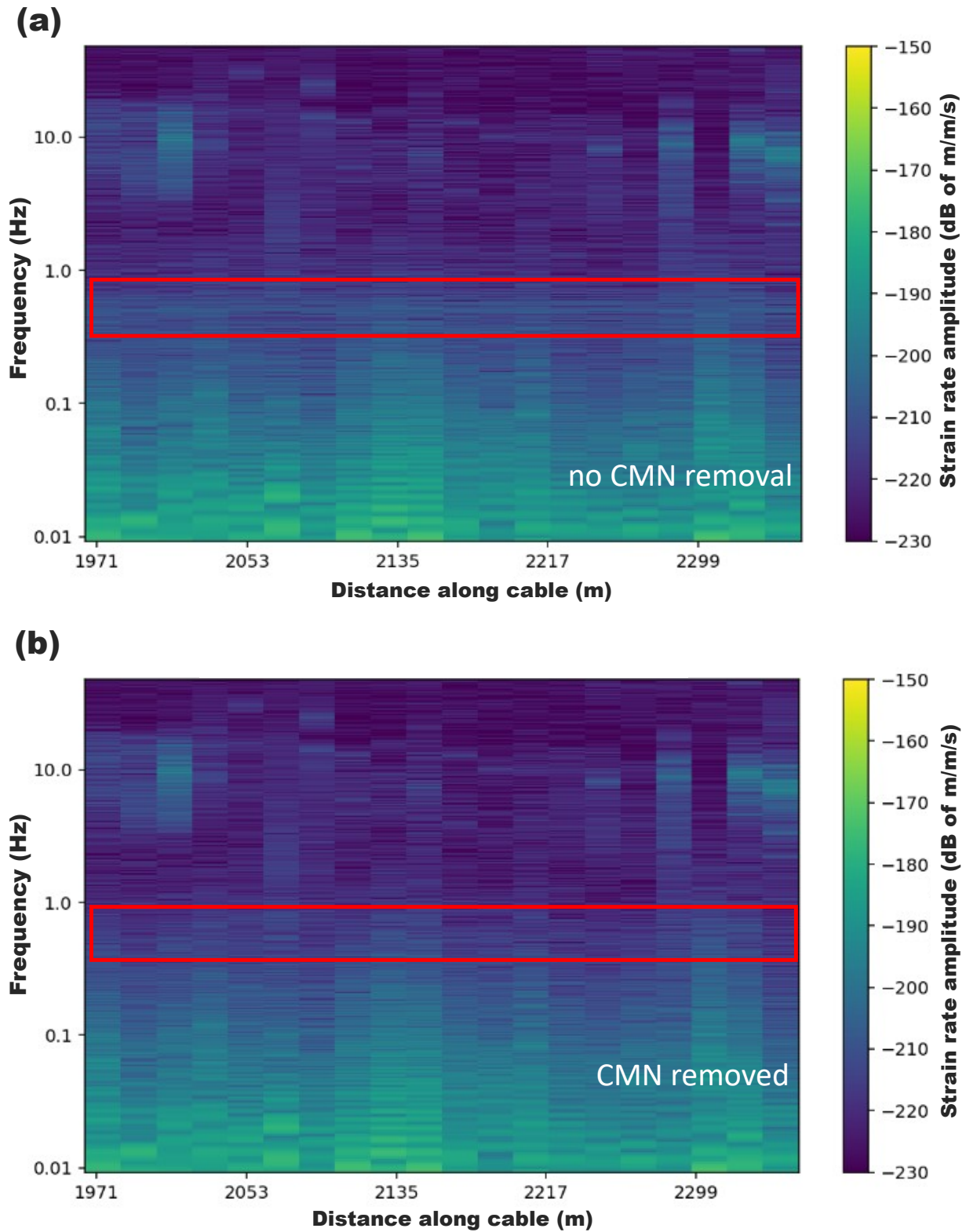

Figure S1: Amplitude spectrum of one day of strain rate DAS data (20/9/2023) for the cable section used for cross-correlation before (a) and after (b) Common Mode Noise removal. CMN is best visible around 0.5 Hz (red box).

Alt-text: Frequency spectra of 20 DAS channels prior and after CMN removal. The CMN, first visible as moderately intense consistent energy around 0.5 Hz, is not visible after the removal.

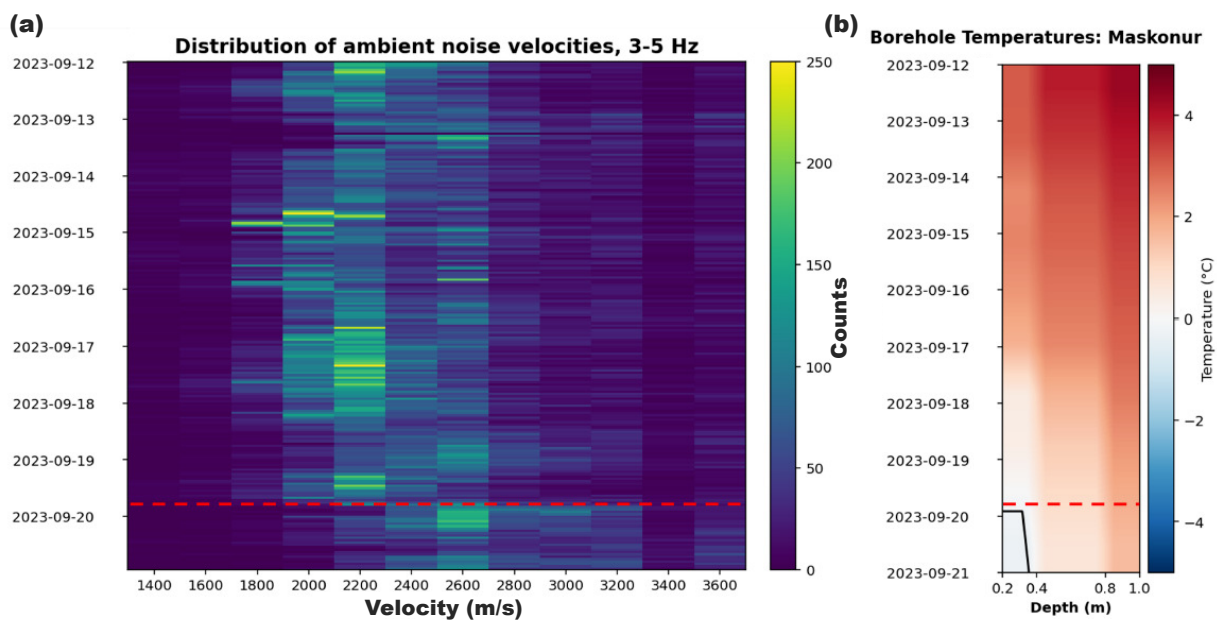

Figure S2: F-K analysis results of ambient seismic noise at HSPA array. (a) Histograms of dominant apparent velocities measured from moving 30 s windows. (b) Evolution of borehole ground temperatures for the same time interval. The red dashed line indicates the beginning of freezing conditions in that area. The velocity increase coincides temporally with ground freezing interpreted from borehole temperature data.

Alt-text: Energy intensity presented as time (2023) vs velocity, showing a shift over time from consistent velocity around 2200 m/s to 2600 m/s. The shift coincides with ground freezing interpreted from temperatures measured in the "Maskonur" borehole.

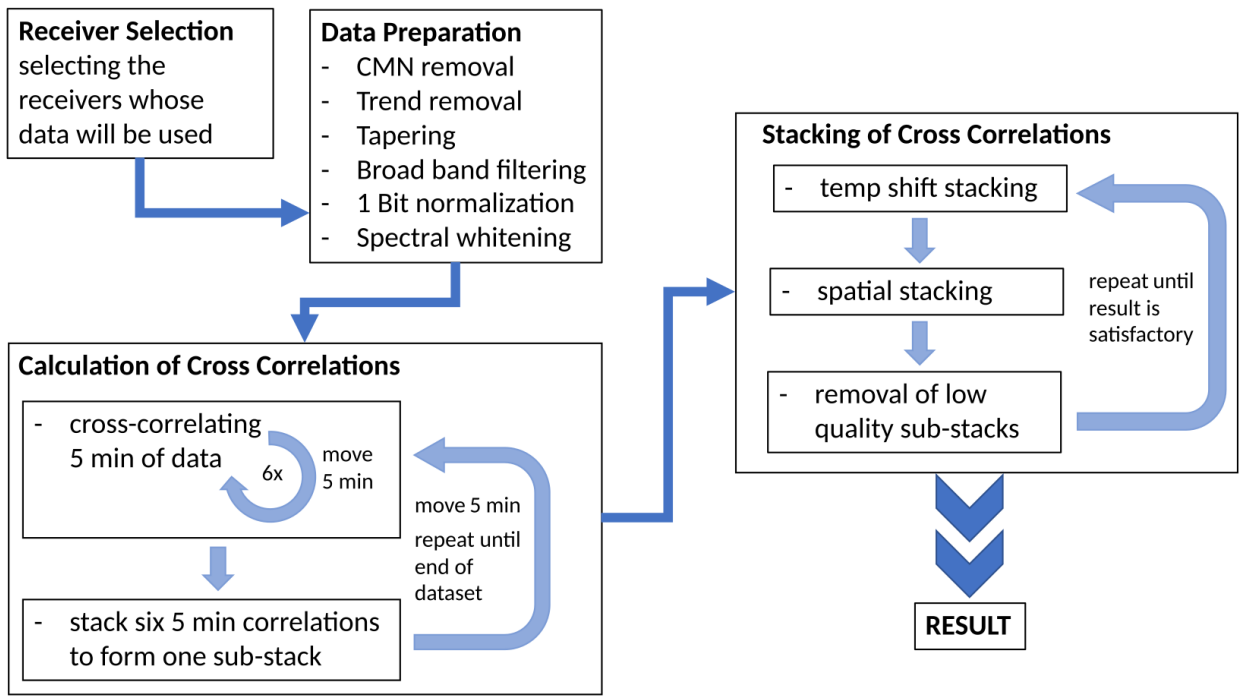

Figure S3: Overview of processing flow used in this study for cross-correlation of DAS data (noise interferometry).
Alt-text: Workflow chart guiding through noise interferometry processing sequence from Receiver Selection via Data Preparation and Calculation of Cross Correlation to final Stacking.

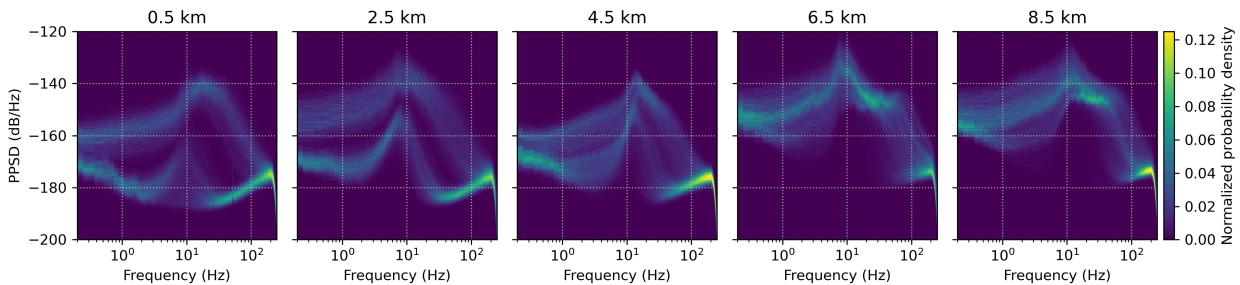

Figure S4: Cumulative strain PPSDs computed every 2 km between 2 and 5 AM between 18 and 21/9/2023 (period covering full fiber span). Tundra section (up to 2.5 km) has a bimodal noise level distribution representing windy days with high noise levels and calm days with lower noise levels. The glacier section (from 6.5 km onwards) is characterized by higher noise levels regardless of the wind speeds. At 4.5 km, where the fiber just enters the ice and is located close to the fjord, a transition between tundra and glacier noise levels regime is visible. Wind speeds are presented in Fig. 2a.
Alt-text: Night-time PPSD computed every 2 km for 2023 data, with the highest noise levels recorded around 10 Hz. Notably, the glacier section exhibits higher noise levels, while those at the tundra section vary between low and high due to wind activity.

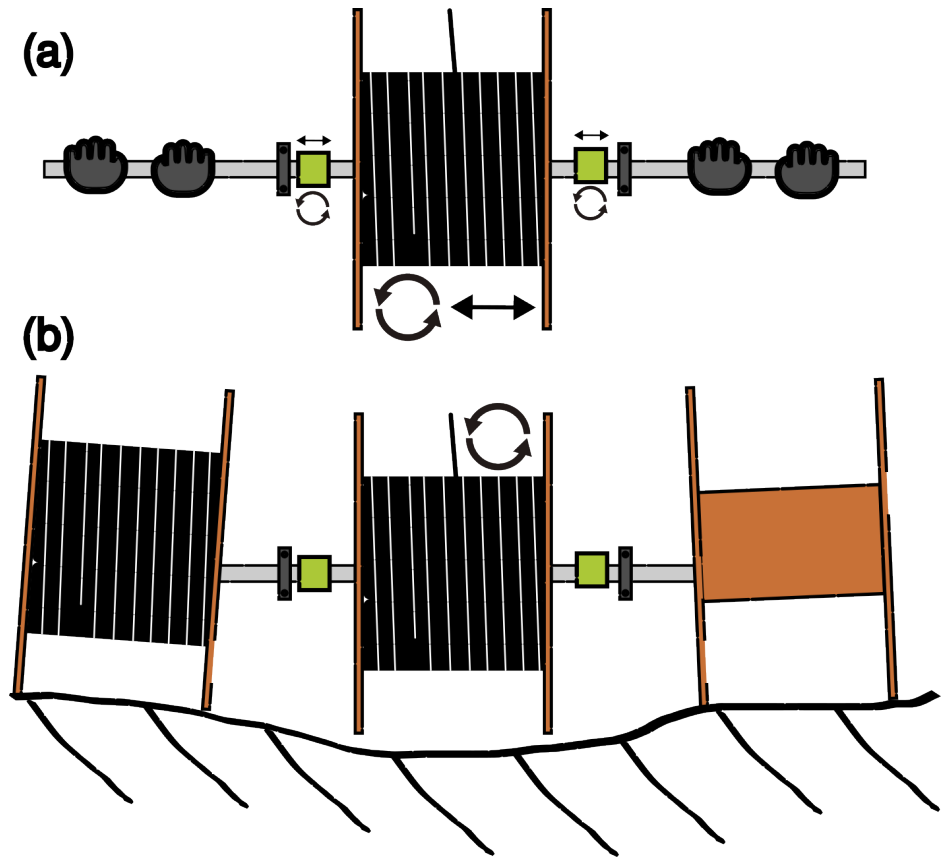

Figure S5: Schematic sketch of the rod adopted for cable deployment and transportation. A 2-m-long sturdy metal rod with fixed elements limiting lateral sliding (gray vertical bars) and mobile rotating plastic segments that facilitate rotating (green boxes), both small enough to allow sliding into the cable drum. Elements that allow sliding and rotation are marked with arrows. (a) The setup used for transporting allowing to unroll the cable while walking. (b) Stationary setup for unrolling equivalent to a standard cable roller. Alt text: Sketch of an improvised field deployment system where a 2-m-long metal rod is used as axis for unrolling the fiber-optic cable from a cable drum. Two cases are shown: unrolling while rod is being held by two people and unrolling from stationary system with metal rod inserted into two additional cable drums.